# Spin-phonon and magnetostriction phenomena in CaMn$_7$O$_{12}$ helimagnet probed by Raman spectroscopy

A. Nonato[a,b,c], B. A. Souza[d], A. P. Ayala[d], A. P. Maciel[e], S. Y. Villar[c], M. S. Andujar[c], A. Senaris-Rodrigues[c] and C. W. A. Paschoal[a]

[a]Departamento de Física, Universidade Federal do Maranhão, Campus do Bacanga, 65085-580, São Luis-MA, Brazil.

[b]Curso Ciências Naturais, Universidade Federal do Maranhão, Campus Universitário de Grajaú, 65940-000, Grajaú-MA, Brazil.

[c]Departamento de Química Fundamental, Universidad de A Coruña, 15071 A Coruña, Spain

[d]Departamento de Física, Universidade Federal do Ceará, Campus do Pici, PO Box 6030, 60455-970, Fortaleza - CE, Brazil

[e]Departamento de Química, Universidade Federal do Maranhão, Campus do Bacanga, 65085-580, São Luis-MA, Brazil.

## Abstract

In this letter we investigated the temperature-dependent Raman spectra of CaMn$_7$O$_{12}$ helimagnet from room temperature down to 10 K. The temperature dependence of the Raman mode parameters show remarkable anomalies for both antiferromagnetic and incommensurate transitions that this compound undergoes at low temperatures. The anomalies observed at the magnetic ordering transition indicate a spin-phonon coupling at higher-temperature magnetic transition in this material, while a magnetostrinction effect at the lower-temperature magnetic transition.

The search for strongly coupled magnetism and ferroelectricity in solids recently led to the discovery of a new class of magnetodielectric materials, whose electric polarization is induced by magnetic ordered states[1-4]. In this case, the electric polarization is strongly sensible to applied magnetic fields, being very suitable for electric control of magnetic properties and, consequently, applicable to spintronics and storage devices. Also, it is usually observed a strong enhancement of the dielectric constant[5,6].

Among the magnetically induced ferroelectrics, $CaMn_7O_{12}$ (CMO) is a singular material, showing two magnetic transitions at low temperatures, exhibiting antiferromagnetic orderings at $T_{N1}$ = 90 K and $T_{N2}$ = 50 K. The magnetic ordering induces a giant ferroelectricity in this compound at $T_{N1}$[7], which is parallel to the c axis due to the in-plane helical magnetic structure[8]. As it is usual in compounds with helical magnetic structure, the change in polarization sign induces a chirality reversal of the helical spiral due to the spin-orbit coupling in CMO. However, in this compound the magnitude of the electric polarization is determined by the exchange striction that permits large values of polarization. This behavior is completely new and unusual[9].

This is a consequence of an unconventional incommensurated orbital ordering below 250 K[10]. According to this model, the orbital occupation rotates perpendicularly to the structural propagation direction, and it has a strong effect on the magnetic ordering at 90 K that CMO undergoes, once it enables the helical magnetic structure to be stabilized by Heisenberg exchange and lock to the superstructure at 90 K[10].

Coupling between phonons (lattice) and the magnetic ordering, which is one of the main mechanisms of coupling between electric and magnetic order in oxides, can be observed at the magnetic transitions. In addition an orbital ordering, which takes changes in the polarizabilities, can induce anomalies in the phonons. Thus, both phenomena should imply in anomalies in the phonon spectra parameters. Raman scattering is a powerful probe to observe the coupling between magnetic ordering and lattice[11–17]. However, recent investigations on low-temperature Raman-active phonon properties in CMO have not detected any anomalies in the magnetic transitions in CMO[18]. In this letter we performed a careful longtime and low laser power Raman scattering investigation of polycrystalline CMO at low temperatures from 10 K up to 300 K to probe the existence of couplings between the magnetic order and the lattice, as well as to investigate the inconmmensurate transition at 250 K.

$CaMn_7O_{12}$ polycrystalline materials were synthesized by the Pechini method[19] as published elsewhere[20-23], using $CaCO_3$ (Panreac, >98.6%) and $Mn(NO_3)_2 \cdot H_2O$ (Aldrich, >98%). First, it was dissolved stoichiometric amounts of these metallic salts in a 1 M citric acid aqueous solution and then added the same volume of ethylene glycol. After, we diluted the obtained solution in water (50% v/v). The resulting solution was heated at 200°C until a brown resin was formed, whose organic matter subsequently was decomposed at 400°C. The precursor powders were ground and then heated in air at 800°C/60 h, 900°C/24 h two times and 950°C/24 h two times, respectively, with intermediate grindings and pelletizing. The resultant powders were pressed into pellets and sintered in air 970°C for 60 h. Room-temperature X-ray powder diffraction (XRPD) patterns of all the samples were obtained with a Siemens D-5000 diffractometer using $Cu(K_\alpha)$ = 1.5418 Å radiation. The obtained XRPD data were analyzed by the

Rietveld profile analysis using Fullprof Suite v.2.05[24]. The analysis confirmed the room temperature structure of CMO as being trigonal belonging to the $R\bar{3}$ space group. Raman spectroscopy measurements were performed using a Jobin-Yvon T64000 Triple Spectrometer configured in a backscattering geometry coupled to an Olympus Microscope model BX41 with a 20x achromatic lens. The 514.5 nm line of an Innova Coherent laser operating at 20 mW was used to excite the signal, which was collected in a N2-cooled CCD detector. All slits were set up to achieve a resolution lower than 1 cm$^{-1}$. Low-temperature measurements were performed by using a closed-cycle He cryostat where the temperature was controlled to within 0.1 K. Magnetic properties were studied in a Quantum Design MPMS Squid magnetometer. In this context, zero-field-cooled (ZFC) and field-cooled (FC) magnetic susceptibility data were obtained under a field of 1000 Oe in the temperature range 5 ≤ T(K) ≤ 300.

Figure 1a shows ZFC and FC susceptibilities of CMO at low temperatures, which confirms its magnetic behavior reported in previous works[23,25]. As expected, this compound is paramagnetic above ∼ 90 K (see inset in Figure 1a), its susceptibility shows a small kink at the antiferromagnetic ordering temperature, $T_{N1}$∼ 90 K. Meanwhile at $T_{N2}$∼50 K the susceptibility markedly increases and the FC and ZFC curves split. Also the curve shows a maximum at $T_g$∼ 20 K, which suggests a spin-glass behavior, where the magnetization in the FC curve becomes virtually constant, while in the ZFC curve has a rapid decrease.

Figure 1b shows the Raman spectrum of CMO at 10 K, which agrees very well with that reported for single crystals of this compound at 8K. As CMO does not change its structure from room temperature until low temperatures, this spectrum is representative forthe rhombohedral structure. We observed 9 Raman-active bands at 10 K. At this temperature (up to

440 K) CMO crystallizes in a trigonal structure, which belongs to the $R\bar{3}(S_6^2)$ space group. According to the site occupation in this structure, the factor group analysis predicts 38 optical phonons in CMO at room temperature, which 12 are Raman-active with symmetries that can be decomposed according to the irreducible representations of the factor group as $6A_g \oplus 6E_g$.

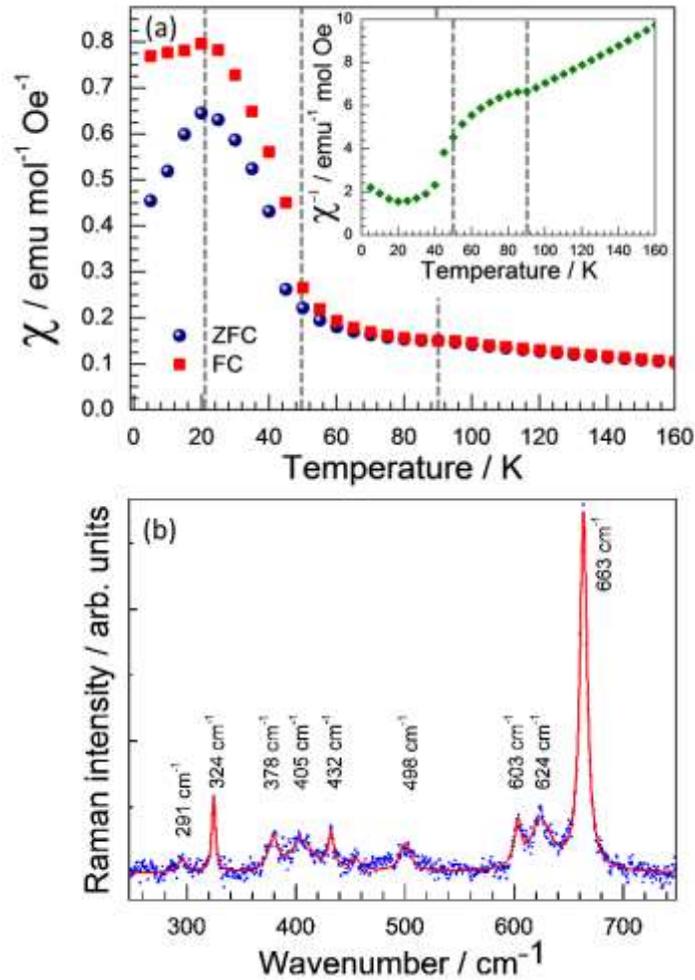

Figure 1 – Top: (a) Magnetic susceptibility of CMO. The inset shows the inverse of magnetization. The dashed gray lines indicate the temperatures in which anomalies were observed. Bottom: (b) Raman spectrum observed for CMO at 10 K.

The observed bands can be assigned according to the careful lattice dynamics performed by Iliev et al [18]. Thus, the phonon observed at 663 cm$^{-1}$, which is associated to the MnO$_6$ stretching mode has symmetry $E_g$, as well as the phonons observed at 603, 405 and 324 cm$^{-1}$. Meanwhile the phonons observed at 624, 498, 458 and 432 cm$^{-1}$ have $A_g$ symmetry. Also, we observed two other bands at 378 and 291 cm$^{-1}$ which are related by symmetry rule breaking due to the incommensurate transition[18].

Usually, the temperature dependence of a phonon position follows the Balkanski model[26], in which the anharmonic contribution to the temperature dependence of the phonon position is given by

$$\omega(T) = \omega_o - C\left[1 + \frac{2}{(e^{\hbar\omega_o/2k_BT} - 1)}\right]$$

with $C$ and $\omega_o$ being fitting parameters. Thus, in the lack of structural phase transitions or any other anomaly, the usual temperature behavior of the phonon position is to exhibit a plateau at very low temperatures; after the position decreases monotonically, exhibiting a negative linear behavior at high temperatures. Figure 2 shows the temperature dependence of the position of some observed phonons. Clearly, the phonons do not exhibit the usual behavior proposed by Balkanski model, exhibiting at least one anomaly in both magnetic anomalies. In the absence of structural phase transitions, this behavior is attributed to spin-phonon coupling, which can be a direct coupling or a coupling due to the magnetostrinction effect[29].

In the first magnetic transition at $T_{N1} \sim 90$ K the anomalies are due to the coupling directly between the magnetic ordering and lattice (phonon) since CMO does not show any anomaly in the lattice parameters in this magnetic transition[23], what discards the possibility of the coupling be due to the magnetostriction effect[27]. However, the second spin-phonon coupling anomaly at $T_{N2} \sim 50$ K could be due to magnetostriction effect, as once at this magnetic transition the lattice parameters of CMO change significantly. In this case, the coupling can be mediated by the magnetostrinction effect[27,28]. To ascertain this assumption, it is more appropriated to investigate the full width at half maximun FWHM (explain the acronym) of the phonons instead of the position, since FWHM is due to the phonon lifetime, which is not affected by magnetostrinction. Thus, FWHM is better to observe if the anomaly is due a direct spin-phonon coupling or induced by magnetostriction[27]. Figure 3 shows the temperature dependence of the FWHM of the stretching phonon. According to Balkanski's model [26], for purely anharmonic contributions, the temperature dependence of a phonon FWHM is given by

$$\Gamma(T) = \Gamma_o \left[1 + \frac{2}{(e^{\hbar \omega_o / 2 k_B T} - 1)}\right] \qquad (2)$$

where $\Gamma_o$ is a fitting parameter. Thus, at low temperatures $\Gamma(T)$ must be constant and monotonically increase when temperature is increased. We can see that $\Gamma(T)$ plotted in Figure 3 shows only a small fluctuation around $T_{N2} \sim 50$ K and an abrupt change at $T_N = 90$ K. Thus, the coupling at $T_{N2} \sim 50$ K is probably due to the magnetostriction effect once is not observed clearly an anomaly at this temperature.

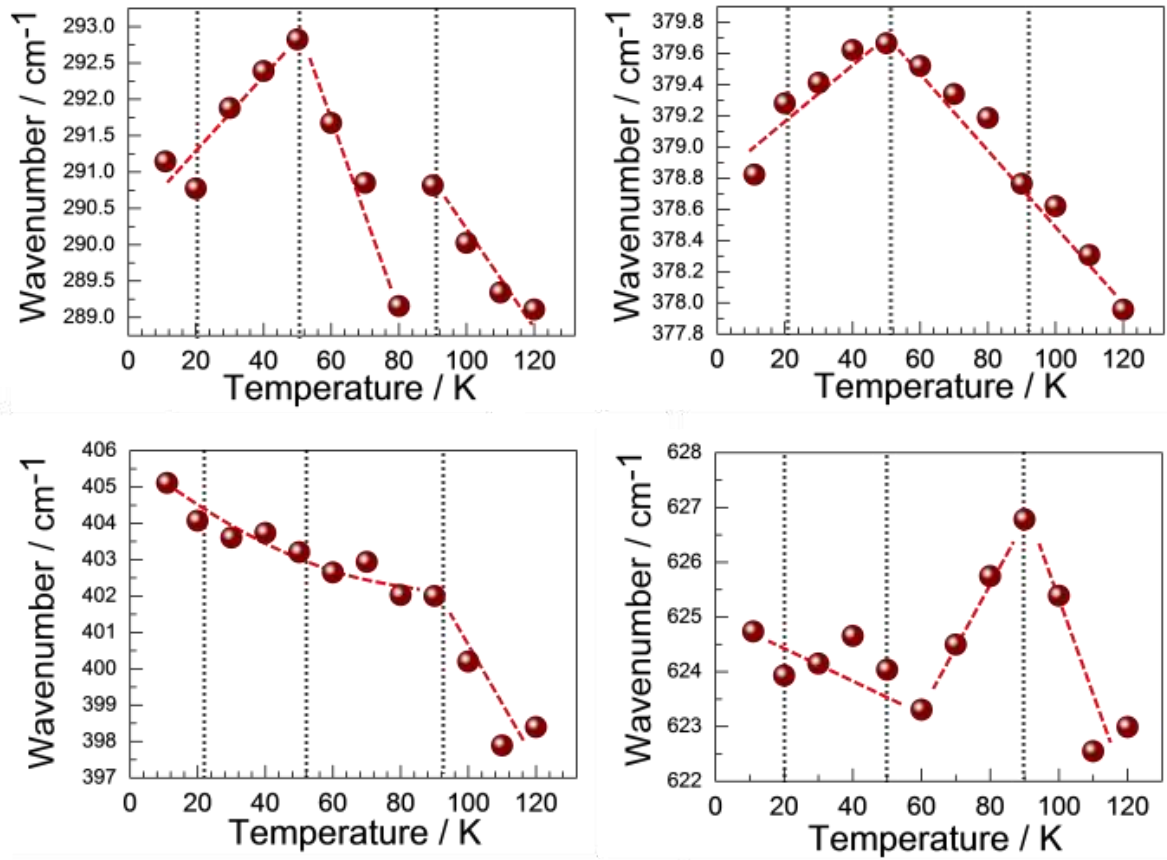

Figure 2 – Temperature dependence of phonon positions up to 120 K. The grey dashed lines indicate the temperatures where the magnetic anomalies where observed, while the red dashed lines are a guide for the eyes.

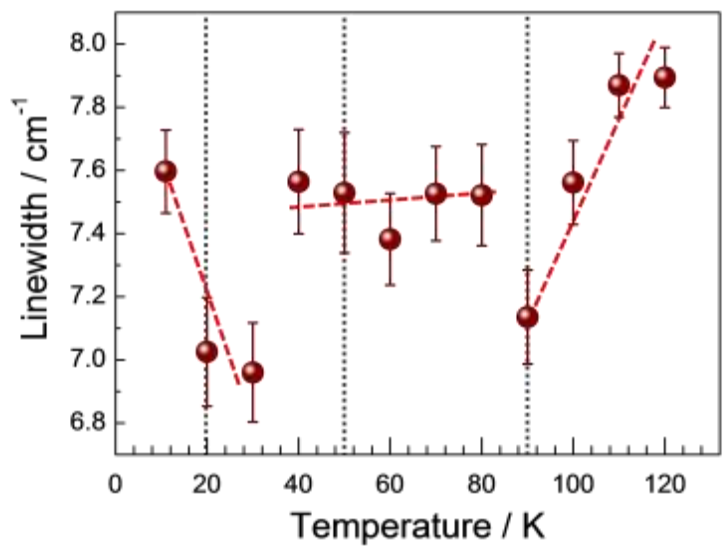

Figure 3 – Temperature dependence of the FWHM of the stretching mode in the temperature range where the magnetic transitions are observed in CMO. The grey dashed lines indicate the temperatures where the magnetic anomalies where observed, while the red dashed lines are a guide for the eyes.

Also, CMO undergoes an incommensurate transition that leads to an incommensurated orbital ordering below 250 K [10]. According to Perks et al[10] in this transition the modulation in Mn-O bond is higher in basal plane of the octahedron, while the modulation in z axis is very small. Thus, we expected that any phonon whose vibration contain atoms in the octahedron basal plane should be susceptible to changes due to the incommensurate transition that originates the helical orbital modulation in CMO at 250 K. Figure 4a shows the temperature-dependence of $MnO_6$ stretching wavenumber position of CMO around the temperature, where

CMO undergoes the incommensurate phase transition (indicated by the gray dashed vertical line). This phonon is due to the $MnO_6$ octahedron and it should exhibit an anomaly at the incommensurate transition due to the orbital/lattice modulation. The stretching phonon position shows only a small deviation from the Balkanski (dotted line) model at approximately 250 K (see solid line), being not conclusive. However, this transition is clearly observed in the normalized Raman intensity of this mode with relation to the phonon at 624 $cm^{-1}$, where it increases abruptly at the transition. This behavior is followed by the normalized intensity of other modes, as indicated in Figure 4b. Thus, assuredly the Raman spectrum is sensible to the orbital ordering at 250 K originated from the incommensurate transition.

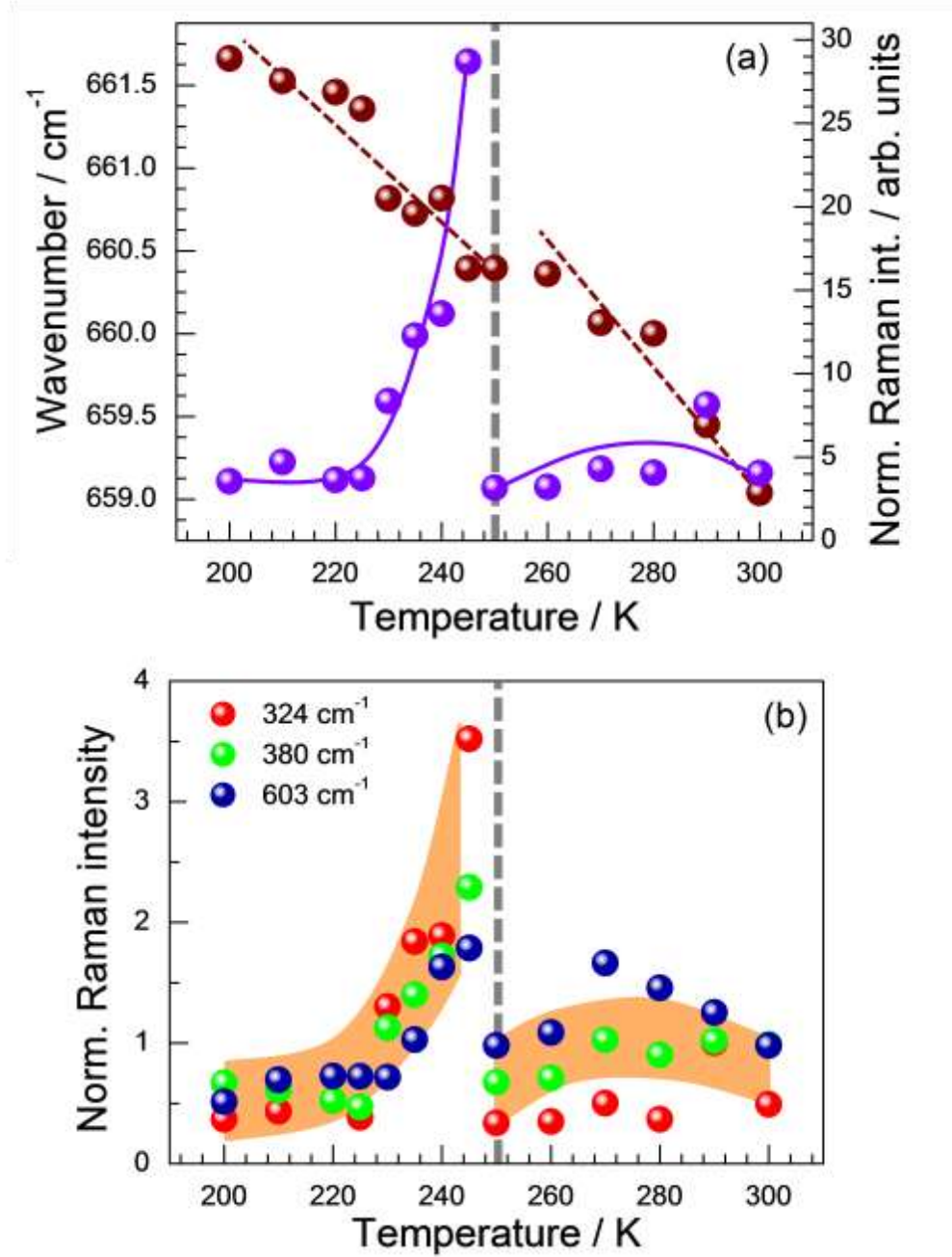

Figure 4 – Top: (a) Temperature dependence of the stretching position mode of CMO (dark red spheres) and its normalized intensity in relation to the phonon at 624 cm$^{-1}$ (purple sphere). The dotted red line indicates the anharmonic contribution to the temperature dependence of the phonon position fitted by Balkanski model, while the solid lines are a guide for the eyes. Bottom: (b) Temperature dependence of the intensity of other three phonons. The shadow is a guide for the eyes.

In this letter we investigated the low-temperature dependent Raman spectra of CaMn$_7$O$_{12}$ ceramics. Both antiferromagnetic transitions at low temperatures, namely 50 K and 90 K, induce anomalies in the phonons positions. The anomalies were associated with spin-phonon coupling and magnetostrinction effects. The incommensurate phase transition at 250 K was observed in the position of the stretching phonon as well as in the intensity of several phonons, which show an abrupt anomaly at the transition.

## Acknowledgments

The Brazilian authors acknowledge the partial financial support from CNPq, CAPES, FUNCAP and FAPEMA Brazilian funding agencies and the Spanish authors are grateful for financial support from Ministerio de Economía y Competitividad (MINECO) (Spain) and EU under project FEDER MAT2010-21342-C02.